\documentclass[a4paper,12pt]{article}
\usepackage{amsfonts}
\usepackage{amsmath}
\usepackage{amssymb}
\usepackage{amsthm}
\usepackage{latexsym}
\usepackage[unicode]{hyperref}
\usepackage{mathtext}
\usepackage[T2A]{fontenc}
\usepackage[cp1251]{inputenc}
\usepackage[english]{babel}
\usepackage{textcomp}
\usepackage{graphicx}
\usepackage{indentfirst}
\usepackage{amstext}
\usepackage{lscape}
\usepackage{makecell}
\usepackage{multirow}
\usepackage{ulem}
\usepackage{mathenv}
\usepackage{indentfirst}
\usepackage{subfig}
\usepackage{geometry}
\begin{document}
\title{Supermassive Black Holes and Kinematics\\ of Disc Galaxies}
\author{A. V. Zasov, A. M. Cherepashchuk, and I. Yu. Katkov\\}
\date{\textit{Sternberg Astronomical Institute, Moscow State University, Moscow, Russia}}
\maketitle

\begin{abstract}
The statistical relations between the masses of supermassive black
holes (SMBHs) in disk galaxies and the kinematic properties of their
host galaxies are analyzed. We use the radial velocity profiles for
several galaxies obtained earlier at the 6-m telescope of the
Special Astrophysical Observatory of the Russian Academy of Sciences
parallel with the data  for other galaxies taken from the
literature. We demonstrate that the SMBH masses correlate well with
the velocities of rotation of disks at a fixed distance  $R \approx
1$ kpc ($V1$), which characterize the mean density of the central
region of the galaxy. The SMBH masses correlate appreciably weaker
with the asymptotic velocity at large distances from the center and
with the angular velocity at the optical radius R$_{25}$. We suggest
that the growth of the SMBH occurs inside of the forming
''classical'' bulge during a monolithic collapse of gas in the
central kpc-size region of the protogalaxy. We have also found a
correlation between the SMBH mass and the total (indicative) mass of
the galaxy M$_{25}$ within the optical radius R$_{25}$, which
includes both baryonic and ''dark'' mass. The masses of the nuclear
star clusters in early-type disk galaxies (based on the catalog of
Seth et al.) are also scaled with the dynamical mass M$_{25}$,
whereas the correlations with the luminosity and velocity of
rotation of galaxies are practically absent for them. For a given
M$_{25}$ the masses of the nuclear clusters are, on average, nearly
order of magnitude higher in S0–Sbc galaxies than in late-type
galaxies.
\end{abstract}

\section{INTRODUCTION}

Recent studies of supermassive black holes (SMBHs) in the nuclei of galaxies have
developed along two directions—investigation of the effects of strong gravitation near the
event horizons of SMBHs (see, for example,
\cite{Gierlinski2008,Doelman2008,Broderick2009}) and analyses of SMBH demographics (see,
for example, \cite{Greene2008}).  Demographic studies require analysis of the relationship
between the characteristics of the central SMBHs and the kinematic properties of the host
galaxies. Information about the history and evolution of SMBHs is ''coded'' in the
morphological and kinematic characteristics of the galaxies, making studies of galaxy
kinematics with known central SMBH masses very promising. Studies of spiral and lenticular
galaxies play a special role here, since measurements of the rotational velocities of the
gas and/or stars in the disks provide direct information about the distributions of the
density and angular momentum.

It is well known that the masses of SMBHs are closely correlated with the parameters of
their host spheroidal systems (an elliptical galaxy or the bulge of a disk galaxy), and
comprise a specified fraction of the total mass of the spheroidal component (about
10$^{-3}$), while the role of the disk is not obvious
\cite{Kormendy2001,Ferrarese2002,Ferrarese2005}. Since the rapid growth of a central black
hole occurs in an early period in the history of a galaxy, when the bulge and disk have
just formed in the gravitational well of the massive dark halo, we would expect the
black-hole mass to be correlated not only with the parameters of the bulge, but also with
the halo mass and the rotational velocity at large distances from the center, $V_{FAR}$,
which determines the virial mass of the galaxy. A relationship between the SMBH mass and
the virial mass of the galaxy is predicted in numerical cosmological models
\cite{Ferrarese2002,DiMatteo2003,Ilin2003,Booth2010} (see also references therein),
although the observational data remain controversial.

Known relations between the SMBH masses and general properties of the host galaxies are
based primarily on data for elliptical and lenticular galaxies. Unfortunately, the number
of disk galaxies that have both reliable black-hole masses $M_{BH}$ estimates and detailed
measurements of their kinematic characteristics (circular velocity, stellar-velocity
dispersion, etc.) is rather small. This has forced to use indirect estimates for
demographic studies of SMBHs: inferring the asymptotic rotational velocities of galaxies
from their empirical dependencies on the central stellar-velocity dispersions
\cite{Ferrarese2002}, and estimating the SMBH masses $M_{BH}$ from their connection with
the stellar-velocity dispersions $\sigma$ \cite{Baes2003}. However, in this case the
resulting relationship between the rotational velocity of the disk and $M_{BH}$ can simply
result from correlations between the indirect methods used to determine these two
parameters. In \cite{Zasov2005} we demonstrated that the correlation between $M_{BH}$ and
the asymptotic rotational velocity is appreciably “looser” than the one which was obtained
from the indirect data \cite{Ferrarese2002,Baes2003}. In view of the poor statistic data,
this conclusion needs to be verified.

We limit our consideration here to disk galaxies. In contrast to
elliptical systems, their disks are rotationally supported, so their
velocities of rotation may give a good idea of the mass contained
inside the chosen radius $R$. It enables to estimate the total  mass
within $R$ depending only weakly on the adopted model for the matter
distribution. We were provided with data from a program of spectral
observations obtained on the 6-m telescope of the Special
Astrophysical Observatory of the Russian Academy of Sciences, for
galaxies with the most reliably determined masses for their central
black holes.  The first results of these observations carried out in
2006–2009 were presented in \cite{Cherepashchuk2010}. To increase
the number of objects for our study, we also use here the kinematic
parameters for other disk galaxies with known $M_{BH}$ based on data
taken from the literature.

\section{MASSES OF THE CENTRAL OBJECTS: KEY PROBLEMS}

In recent studies of the formation of SMBHs and the growth of their masses, three
important problems have been prominent.

1. The problem of the very fast growth of the SMBH masses ($M_{BH}$) at high redshifts (at
least for massive galaxies and quasars). This rapid growth is implied by the discovery of
more than ten quasars with very high redshifts $z \approx 6~-~7$ \cite{Volontieri2006,
Mortlock2011}, as well as the gigantic SMBH masses (up to $4 \times 10^{10}$
M$_{\bigodot}$) for some objects at $z \approx 4$ \cite{Ghisellini2009}.  This probably
suggests the direct formation of initial black holes with masses of $10^4–10^5$
M$_{\bigodot}$ in the inner regions of forming galaxies, although this scenario requires a
mechanism that can slow the rapid cooling of the gas and its transformation into stars
(see the diskussion of this problem and references to original studies in
\cite{Mayer2010}). The growth of $M_{BH}$ at high redshifts apparently overtakes the
growth of their bulges, whose masses increase more slowly. This conclusion has emerged
from both numerical calculations of cosmological evolution \cite{Croton2006,Colberg2008}
and direct analysis of observations of quasars \cite{Salviander2007} and Seyfert galaxies
\cite{Woo2006}.

2. The relationship between the central black holes and the nuclear clusters (NCs) remains
unclear; the latter are observed in the centers of both spiral and elliptical galaxies
and, as a rule, have modest luminosities. Both of these formations are often unified under
the term ''central massive object'' (CMO). As a rule, the masses and sizes of the NCs
exceed the most massive globular clusters in the Galaxy, and have more complex
star-formation histories, which are different for different galaxies and do not correspond
to a single star-formation bursts \cite{Marel2007}. The two types of СМО (SMBH and NC) can
exist independently of each other, although several galaxies where they are observed
together are known \cite{Seth2008,Torsten2010}. In all such cases, the black-hole mass
exceeds the mass of the NC.

3. A key problem is to explain the observed correlations between the  mass of the SMBH
and/or NC and the bulge properties such as $M_{BH}~-~\sigma$ dependence. The SMBH masses
statistically depend not only on the velocity dispersion, but also on the structure of the
bulge. It was  found that the masses of the central black holes in pseudobulges are, on
average, a factor of a few lower than in galaxies with classical bulges or E galaxies, for
the same central velocity dispersions \cite{Hu2008}. There may also exist an analogous
difference for the NCs, which are, on average, less massive in late-type galaxies, for a
specified mass of the stellar population of the galaxy \cite{Seth2008}: for the same
stellar-velocity dispersion, $M_{BH}$ is, on average, several times higher for classical
bulges than for pseudobulges \cite{Hu2008}.

The difference in the SMBH masses for galaxies with classical bulges and pseudobulges is
an important fact. Pseudobulges have lower central brightness concentrations, higher
degrees of flattening, and more fast rotation. They are often considered to be false
bulges—the result of ''heating'' of the inner region of the disk (for example, due to the
evolution of an existing or formerly existing bar, or during a redistribution of the
angular momentum of baryonic matter in the disk), while classical bulges are usually
considered to result from mergers in the early stages of evolution of the galaxy.
Pseudobulges are possessed by late-type Sc–Sd galaxies and some earlier-type galaxies with
small bulges, with Sersic parameters $n \leq 2$ \cite{Kormendy2004,Fisher2008}.  Below we
consider a possibility that at least one of the  key parameters determining the growth
rate and final mass of the SMBH in a young galaxy is not the type of bulge, but rather the
mean density of matter (initially gas) in the central, kpc-size part of the observed
bulge. It is the density of matter which determine the gravitational contraction time of
gas, the rate at which the gas is transformed into stars, and the accretion rate onto the
CMO.

\section{RELATIONSHIP BETWEEN THE VELOCITY OF ROTATION OF THE DISC AND MASS OF CMO}

\subsection{Relationship between $M_{BH}$ and the Angular Velocity of the Central Region}

The mean density of the inner region of a galaxy within a radius $R$ is proportional to
the square of the angular  velocity $V/R$ at this distance. Therefore, it makes sense to
test for the existence of a universal (for bulges and pseudobulges) correlation between
the SMBH mass and the angular velocity, for which we will use the circular velocity at a
fixed distance $R = R_b$. The choice of $R_b$ is fairly arbitrary. Here, we choose the
velocity $V1$ corresponding to $R_b = 1$ kpc. First, this is the characteristic size of
the dynamically and/or chemically decoupled nuclear regions of disk galaxies. Second, as a
rule, the finite angular resolution of observations hinders determination of the shape of
the rotation curve with the desired accuracy at distances closer to the center. Third,
noncircular motions of gas associated with peculiarities of the inner structures of
galaxies (bars, spirals, rings, inclined disks, active nuclei) are frequently observed
within the central kiloparsec. Note that the first attempts to identify and study the
kinematic properties of the inner kpc-size regions in galaxies with active nuclei were
undertaken by Afanas’ev, who plotted the rotational velocity of the gas against the mean
volume luminosity of the bulge at this distance \cite{Afanasev1987,Afanasev1986}.

Figure 1a compares the masses of the central black holes $M_{BH}$ and the rotational
velocities V1 of parent galaxies at $R_b \approx$ 1 kpc based on observations obtained at
the SAO 6-m telescope \cite{Cherepashchuk2010}. The two objects are added here: 3C~120,
whose rotational velocity we measured with lower accuracy due to the bright nucleus, and
the massive S0 galaxy NGC~524, whose data were reduced later. The inclination of NGC~524
derived from several spectral cuts is found to be about 35$^{\circ}$. For the most slowly
rotating galaxy (NGC~428), we used the mass of the nuclear star cluster (taken from the
data pr5esented by Seth et al. \cite{Seth2008}) instead of the black-hole mass which is
unknown. The relationship is preserved, but appears looser if to replace V1 onto the
velocity of rotation at a maximal  distance from the center $V_{FAR}$ where it is
measured, or asymptotic velocity (Fig. 1b). However, this conclusion is not too convincing
because for the galaxies considered here the velocities at large $R$ are estimated with
larger uncertainties than the velocities at a fixed distance.

To increase the statistics, we used published data on the rotation curves of galaxies with
known blackhole masses. Figure 2 presents the diagram analogous to Fig. 1a for an
appreciably larger number of objects (see Table), whose black-hole masses have been
estimated using the most reliable methods: reverberation mapping and the resolved
kinematics method, in the optical or the radio ranges (the latter is from observations of
megamasers). We have also included several galaxies with model SMBH masses based on
optical line-of-sight velocity measurements with very high angular resolution (NGC~404,
NGC~524, NGC~3368, NGC~4435). We have only upper limits on the blackhole masses for M33,
NGC~4435, and IC~342. The lowest upper limit of $M_{BH}$ in the diagram is that for М33.
For galaxies with measured masses for both the central black holes and the NCs the latter
are indicated by asterisks. In these cases pairs of corresponding mass values are joined
by vertical lines. The rotational velocities were estimated from the rotation curves cited
in the ''Bibliographical catalog of galaxy kinematics'' in the HYPERLEDA database (see
also references in \cite{Zasov2005}). When several rotation curves were available, we gave
preference to the one that was traced most certainly in the inner region of the galaxy.
The characteristic uncertainties are illustrated by the cross in the lower-right corner of
the figure. The uncertainties in the SMBH masses were taken to be a factor of two (0.3
dex) \cite{McGill2008, Marconi2003} if a larger uncertainty was not indicated in the
source, while the uncertainties in the velocities were taken to be 25\% ($\sim 0.1$ dex).
The empty circles in Fig. 2 denote galaxies with pseudobulges. The latter include
late-type galaxies (Sc and later) and earlier-type galaxies for which the presence of a
pseudobulge is indicated by photometry (following \cite{Hu2008,Fisher2008}).

It follows from Fig. 2 that the black-hole masses are indeed correlate with the angular
velocity of the disk at $R \approx 1$ kpc, at least for galaxies whose velocities of
rotation exceed 200 km/s. After exclusion of M33, the correlation coefficient is $r =$
0.64. The galaxies М33 and IC~342, for which only upper limits for the SMBH masses are
available, are located below the general sequence, but they agree with it, if the mass of
the black hole is replaced by the mass of the NC or the sum of the two masses.

\subsection{Dependence of $M_{BH}$ on Other Kinematic Parameters}

The relationship between the SMBH mass and the rotational velocity becomes less tight if
the latter velocity corresponds to the outer parts of the observed rotation curve
$V_{FAR}$, where a curve of rotation reaches maximum or a plateau (Fig. 3a). Thus, our
conclusions do not support the presence of a tight correlation between the black-hole
masses and circular velocities of galaxies far from the center.

This correlation also nearly disappears when we plot $M_{BH}$ against the angular velocity
of the disk at the optical radius $R = R_{25}$ (where $R_{25}$ is the radius corresponding
to the surface brightness 25$^m$/arcsec$^2$ in the $B$ band) instead of $R = 1$ kpc (Fig.
3b). Since the observed rotation curves of our sample galaxies  reach $R_{25}$ in only a
few cases, the velocities of rotation were taken from the HYPERLEDA database; in most
cases, they were based on the H I linewidth measurements.

The ''classical'' dependence of the masses of the black holes (and of several NCs in the
same galaxies) on the central stellar velocity dispersion $\sigma$ (taken from the
HYPERLEDA database), presented in Fig. 3c. In general, this plot looks rather similar to
the diagram ``$M_{BH}$-V1'' (Fig. 2), although galaxies with pseudobulges (open circles)
are shifted in this case.

Note that although the relation between $M_{BH}$ and $V1$ is due to the same factors as
the dependence of $M_{BH}$ on the central velocity dispersion $\sigma$, the former does
not simply reduce to the latter. First, observational estimates of $\sigma$ are the result
of averaging of chaotic stellar velocities along the line of sight. The velocity
dispersion falls with distance from the center, and the result depends on the adopted
radius for the averaged region. This radius has been determined in different ways in
different studies, and is usually related to the effective radius of a galaxy in some way.
Second, in contrast to the rotational velocity, $\sigma$ is not related to either the mass
or the mean density of the system; translation to the latter requires the construction of
a dynamical model and, in general case, taking into account the rotation of a bulge, which
is usually badly known. Finally, measurement and interpretation of the velocity dispersion
becomes especially complex for galaxies with small bulges, where two dynamically distinct
components contribute to $\sigma$ -- the bulge and the disk. The effect of the disk is
especially important in the presence of a young stellar population with a low velocity
dispersion. At the same time, although the circular velocity of the disk at a given
distance $R$ is often measured with lower accuracy than $\sigma$, the former has a simpler
universal interpretation, and in all cases characterizes the total mass of matter within
the chosen radius.

\subsection{Relation between $M_{BH}$ and the Luminosity and Mass of the Galaxy}

Figure 4 plots the mass of the central black hole against the total luminosity $L_V$ and
the indicative mass $M_{25} = V^2_{FAR}R_{25}/G$, which is close to the total mass of the
galaxy within the optical radius $R_{25}$. As expected, the mass of the CMOs is very
loosely connected with the total luminosity of galaxy. Indeed,  black-hole mass is known
to closely tied with the mass and luminosity  of the spheroidal component of a galaxy only
\cite{Kormendy2001,Marconi2003} (see also references therein). However, there is an
overall trend for all galaxies to increase $M_{BH}$ with total mass $M_{25}$ (the
correlation coefficient for the logarithmic plot is $r = 0.61$). Note that a bulge mass
contribution to $M_{25}$ is usually small.  The linear regression fit in Fig. 4 has the
form

\begin{equation}
\log M_{BH} = a \log M_{25} + b,
\end{equation}

where $a = -9.0 \pm 3.4$, $b = 1.50 \pm 0.31$ (without including the galaxies with upper
limits for $M_{BH}$).

In contrast to the total luminosity of the galaxy, which is closely correlated with its
rotational velocity (the Tully–Fisher relation), $M_{25}$ depends on the masses of both
the baryonic components (stars, gas) and the dark halo, with their contributions being
comparable within the optical radius (see, for example, \cite{Zasov2011}). Ferrarese et
al. \cite{Ferrarese2006} show the existence of a similar dependence between the mass of
the CMO and the quantity $M_{gal} \sim R_e \sigma^2_e/G$, where $\sigma_e$ is the
stellar-velocity dispersion within the effective radius $R_e$, which contains half the
total luminosity of an elliptical galaxy or the bulge of a spiral galaxy. However, their
conclusions of \cite{Ferrarese2006} were based mainly on data for elliptical galaxies.

\section{CMOs IN GALAXIES WITH CLASSICAL BULGES AND PSEUDOBULGES}

 In contrast to early-type galaxies, late-type Sbc–Sd galaxies always possess
pseudobulges, while, as a rule, earlier type galaxies contain classical bulges or both
types of bulges. Therefore, the division of galaxies into early and late morphological
types can be viewed as a division into objects with classical bulges and pseudobulges
(although this identification is not perfect). Hu \cite{Hu2008} noted that the SMBH masses
in galaxies with pseudobulges (i.e., in late-type galaxies) are systematically lower than
those in early-type galaxies, for the same velocity dispersion. This is illustrated for
our sample of galaxies in Fig. 3c, where we compare the masses of black holes and several
nuclear clusters with the central velocity dispersion taken from the HYPERLEDA database.
However, galaxies with the two types of bulges are well mixed in the diagram where
$M_{BH}$ is plotted vs. $V1$ (Fig. 2). This suggests that, for a given mean density of
matter within the central kpc-size region, $M_{BH}$, or more generally the CMO mass, does
not depend, or depends only weakly, on the type of a bulge.

The difference between galaxies with bulges and pseudobulges, or between galaxies of early
and late types, is especially clear when considering the stellar masses of the NCs, which
was demonstrated by Seth et al. \cite{Seth2008}: nuclear clusters in the early type
galaxies are more massive. As in the case of the blackholes, there is essentially no
correlation between the masses of nuclear clusters and the velocities of rotation of their
parent galaxies. This is illustrated by Fig. 5a, where the masses of NCs are plotted
against the velocities of rotation of galaxies taken from HYPERLEDA (they are obtained
mainly from the HI linewidths). Galaxies with disk inclinations $i < 30^{\circ}$ were
excluded due to uncertainty in the estimated corrections for the disk inclination. The
empty symbols indicate galaxies of type Sc or later (i.e., galaxies with pseudobulges or
without any appreciable bulges). However, as in the case of the SMBHs, there is a
correlation between the NC masses and the total dynamical masses of the galaxies within
the optical radius $M_{25}$ (Fig. 5b), which is especially clearly manifests for S0–Sbc
galaxies (correlation coefficient $r = 0.65$). The correlation with galaxy mass is weak or
absent for late-type galaxies possessing only small (pseudo)bulges (empty circles). The
masses of the NCs are, on average, a factor of six lower than in the earlier-type
galaxies.

Earlier, Seth et al. \cite{Seth2008} found a similar relation between the NC mass and the
total mass of stellar population of a galaxy, estimated from its luminosity (in the $B$
band) and color index (see Fig. 2 in \cite{Seth2008}). The mass values found from
photometry and on the base of circular velocities  are not identical, since the
kinematically found  mass is the sum of masses of visible components and a dark halo,
while the photometric mass relates only to stellar population, being calculated under
certain assumptions about the initial stellar mass function and the star-formation
history. It is striking that the difference between early- and late-type galaxies in Fig.
5b reveals itself appreciably clearer than the similar relation based on the masses of
stellar population alone presented by Seth et al. \cite{Seth2008}.

To resume, both SMBHs and NCs masses depend weakly on the linear or angular velocities of
rotation of the disks at large radii, but they correlated with the angular velocity within
the central kiloparsec, and also with the mass of a galaxy within the optical radius.  We
have confirmed that galaxies with pseudobulges, which are primarily late-type spiral
galaxies, have appreciably less massive NCs, on average, for the same integrated
characteristics of the galaxy and the same stellar-velocity dispersion
\cite{Ferrarese2006}.

\section{DISCUSSION}

The genetic relationship between the central black-holes and stellar clusters is obvious:
their masses are correlated with the mass or stellar-velocity dispersion of the bulge in
the same way. However, the attempts to explain how the formation of one could be connected
with the presence of the other encounter considerable difficulties. The ratio of the NC
and SMBH masses encompasses a very large range. As a rule, in galaxies with low-luminosity
bulges, $M_{BH} < M_{NC}$ \cite{Ferrarese2006}, while the opposite is true for large
bulges \cite{Graham2009}. There also exist galaxies with high luminosity in which there
exist a SMBH, but no  NC is observed. Partially it can be explained by the difficulties in
detecting star clusters against the bright background of the nucleus of a massive galaxy
(although it is hard to ''hide'' a star cluster with a mass of $10^9$M$_{\bigodot}$), but
this is more likely associated with an early cessation to the growth of NCs in galaxies
with the most massive black holes. A key factor here is the difference in the conditions
for the growth of the masses of SMBHs and NCs.

The correlation between the masses of SMBHs and NCs with the properties of bulges rather
than the disks of their parent galaxies suggest that both types of CMOs arise from the
same gas medium as the stars in the central part of bulges, but via different processes
that occur on different time scales. The depth of the potential well at the galactic
center is of primary importance here. Classical bulges and pseudobulges probably formed at
different times (the latter as a product of the evolution of the inner part of the disk).
Pseudobulges rotate faster than the classical ones, and, since galaxies with this type of
bulges are not distinguished in a $M_{BH}-V1$ diagram, we conclude that the final mass of
the central black hole is more closely related to the matter density in the inner kpc-size
region than to the specific angular momentum of the bulge.

The inner region of a galaxy containing the densest part of the bulge, with which we
associate the growth of the SMBH, should form within the first billion years, in the first
and relatively short stage of the galaxy formation. The presence of two stages in the
formation of galaxies has been investigated in a number of studies numerically modeling
this process in a $\Lambda$CDM model of the Universe (see, for example,
\cite{Cook2009,Xu2007,Wechsler2002}). According to the numerical simulations of Cook et
al. \cite{Cook2009}, during the first, short phase in the formation of galaxies, there was
a collapse of dark and baryonic matter into the inner regions of the galaxies, leading to
the formation of stellar spheroids (bulges) at $z \approx 2$, after which there began a
quieter and more prolonged stage of disk formation, which did not influence the galactic
centers. In the simple analytical model proposed by Xu et al. \cite{Xu2007}, the stellar
bulge forms as a result of a gravitational monolithic compression of an isothermal gaseous
sphere, which loses its stability in the gravitational field of the central cusp of the
dark halo; this process is accompanied by violent star formation, which sustains the
growth of the central black hole.

Evidently, the formation of both the bulges of disk galaxies and the elliptical (E)
galaxies is a complex process. Baes et al. \cite{Baes2007} separated the influence of the
age and chemical composition on the spectrum of the stellar population for  elliptical
galaxies, and concluded that the star-formation histories are different for the inner and
outer regions of a galaxy: the central region ($R \approx 1.5–2$ kpc) has a steep
metallicity gradient, and apparently formed as a result of a monolithic collapse of gas,
in contrast to more distant regions. The hydrodynamical computations of Pipino et al.
\cite{Pipino2010} subsequently demonstrated the agreement between the observed metallicity
gradients in E galaxies and those expected for a monolithic collapse.

Classical bulges of disk galaxies are similar to E galaxies: they are characterized by the
similar relations between the observable parameters \cite{Fisher2008}. So one can expect
that the inner regions of disk galaxies with the size of the order of a kiloparsec,
containing the classical bulge, could also have formed from gas compressed as a whole,
without a participation of merging of smaller subsystems. A monolithic collapse of gas is
natural to associate with the fast growth of the CMOs (SMBHs and/or NCs). As
three-dimensional hydrodynamical computations show, the higher the gas density, the more
intense the accretion of gas into the center of the forming galaxy and, as a consequence,
the more massive the central object that is formed—a future SMBH \cite{Li2007}.

A high gas density in the inner part of a protogalaxy leads to intense star formation near
the galactic center. On the one hand, this process facilitates the growth of the black
hole, since it provides a source of turbulent motions in the gaseous medium (see, for
example, \cite{Escala2006}); on the other hand, it limits the growth of the black hole in
time, since it depletes the supply of available gas \cite{Li2007}. If the bulge has a low
density and star formation occurs over an extended time, the central black hole is not
able to grow effectively, and it remains to be a comparatively low-mass object. However,
the growth of the bulge mass can continue for a long time after the growth of the
black-hole has ceased — for example, as a consequence of mergers with small systems, or as
a result of dynamical heating of the inner region of the disk (in this case a pseudobulge
may  form). Numerical simulations of the growth of a bulge show that this scenario can
explain the gradual emergence of the ratio between the bulge and SMBH masses
\cite{Croton2006}. Note that in galaxies where both types of bulges are observed (a more
compact classical bulge and a pseudobulge) the mass of the black hole  correlates with the
mass of the classical bulge \cite{Erwin2010}.

The formation of a galaxy took place in the gravitational field of both dark halo and
baryonic matter. The role of dark matter in formation of CMOs is poorly known. In
generally accepted cosmological schemes for galaxy formation, the virial mass of a galaxy,
which is comprised primarily by the dark halo, is determined by the value of circular
velocity at large $R$, which is close to the observed velocity $V_{FAR}$
\cite{Dutton2010}. Nevertheless, as it was shown above, the masses of CMOs correlate
better with $M_{25}$ than with $V_{FAR}$. It is worth reminding that $M_{25}$ is the total
mass within the optical radius of a galaxy, where the masses of baryonic and dark matter
are usually comparable.

Both types of compact objects -- SMBH and NC -- may begin to grow independently,
increasing their masses more rapidly if the gas density in the central part of forming
galaxy is higher.The correlation between the masses of nuclear clusters and bulges is not
as tight as it is for $M_{BH}$ (see \cite{Erwin2010}), which may reflect that the growth
stage of the cluster lasts appreciably longer than the growth stage of the black hole.
Nuclear clusters in E galaxies can increase their masses via the accretion of gas with low
angular momentum, and those in spirals -- via the slow accretion of gas from the disk, or
due to dynamical friction of massive objects in the disk (star clusters,massive gas
clouds) \cite{Agarwal2010,Bekki2009,Capuzzo2008}. Therefore, when the central black hole
has only a modest mass (say, due to a low initial gas density), the NC can overtake the
black hole in mass over billions of years, while still remaining comparatively low-mass
(as in M33, for example). In galaxies with massive classical bulges, and consequently with
high-mass central black holes, a massive NC may not form, due to the expected high
activity of the nucleus; this apparently is the case in early-type high-luminosity
galaxies.

\section{CONCLUSION}

The main conclusions of this work are the following.

1. The masses of supermassive black holes (SMBHs) in the nuclei of disk galaxies do not
depend on the angular velocity of peripheral regions of the disk, but they correlate with
the angular velocity of the inner kpc-size regions of galaxies, which characterizes the
mean density of the matter there. Galaxies with classical bulges and pseudobulges probably
form a single sequence, although better statistics are needed before firm conclusions can
be drawn. Based on the idea of Baes et al. \cite{Baes2007} and the model of Xu et al.
\cite{Xu2007}, which propose that there should be a monolithic collapse of matter during
the formation of the inner regions of galaxies, we suggest that such a collapse of
pseudo-isothermal gas within a radius of $R \approx 1$ kpc is responsible for the rapid
growth of mass of the central SMBH and the classical bulge in the initial period of a
history of galaxy.

2. Although a correlation between the SMBH mass $M_{BH}$ and the circular velocity
$V_{FAR}$ at large distances R can be traced, it is fairly loose, confirming our earlier
conclusion \cite{Zasov2005}. Consequently, the velocity of rotation of the outer disk,
which characterizes the total virial mass of the dark halo in modern models for galaxy
formation, does not play the determining role in the formation of the central black hole.
The same may also be true for nuclear star clusters (NCs).

3. The masses of both the SMBHs and NCs are correlated with $M_{25} = V^2_{far}R_{25}/G$,
where  $M_{25}$ is the indicative total mass of a galaxy within the optical radius
$R_{25}$. A similar relation for elliptical galaxies, whose masses were crudely estimated
based on their velocity dispersions, was demonstrated earlier by Ferrarese et al.
\cite{Ferrarese2006}. This enables to suggest that the conditions for the formation of the
central massive objects in both disk and E galaxies depend on the mass of the galaxy
within the optical radius, where the contributions of dark and baryonic matter are
comparable, rather than with the mass of a dark halo only.

4. Nuclear stellar clusters in early-type spiral galaxies (S0–Sbc) are, on average, nearly
an order of magnitude more massive than those in later-type galaxies with the same values
of $M_{25}$, or than the SMBHs in these galaxies. It concords with the idea that the
formation of SMBHs and NCs occurs on different time scales, and the masses of the NCs (if
they managed to form close to the black hole) apparently continue to grow after the growth
of the SMBHs has ceased.

5. At least for spiral galaxies with central black holes with comparatively modest masses,
the total mass of the NC$+$SMBH correlates better with such parameters as the central
angular velocity of a disk and the indicative mass $M_{25}$, than does the mass of  SMBH
alone.

\section{ACKNOWLEDGMENTS}

The authors thank A.V. Moiseev and V.L. Afanas’ev for obtaining the spectral observations
on the SAO 6-m telescope, and also O.K. Sil’chenko for interest in this work and valuable
comments.

\pagebreak

\begin{table}
\caption{Black-hole (central cluster) masses and kinematic parameters for the galaxies
considered}
{\scriptsize
\begin{tabular}{|c|c|c|c|c|c|c|}
\hline
 \textnumero & Galaxy    & $M_{BH,NUCL},10^6 M_\odot$        &  Source  & V1,km/s  & $V_{FAR}$,km/s & $\sigma$,km/s \\
\hline
  1 & NGC 224   & 140   & \cite{Graham2008}       & 230 & 232 & 170\\
  2 & NGC 404 (BH)  & 0.45  & \cite{Seth2008}       & - & 200 & -\\
  3 & NGC 404 (NUCL)& 10    & \cite{Seth2008}       & - & 200 & -\\
  4 & NGC 598 (BH)  & <0.0015 & \cite{Gebhardt2001}   & 60  & 130 & 37\\
  5 & NGC 598 (NUCL)& 2 & \cite{Kormendy1993}   & 60  & 130 & 37\\
  6 & NGC 1023  & 44    & \cite{Ferrarese2005} & 130 &  -  & 204\\
  7 & NGC 1068  & 8.3   & \cite{Ferrarese2005,Jones1999}  & 220 & 230 & 199\\
  8 & NGC 1300  & 66    & \cite{Atkinson2005} & 170 & 230 & 229\\
  9 & NGC 2748  & 44    & \cite{Atkinson2005}   & 115 & 145 & -\\
  10 & NGC 3031 & 70    & \cite{Ferrarese2005}     & 300 & 168 & 161\\
  11 & NGC 3227 & 7.63  & \cite{Denney2010}       & 140 &   - & 133\\
  12 & NGC 3368 & 7.5   & \cite{Nowak2010}      & 160 & 200 & 128\\
  13 & NGC 3384 & 16    & \cite{Ferrarese2005} & 106 & 200 & 148\\
  14 & NGC 3783 & 29.8  & \cite{Peterson2004} & 150 & 180 & 155\\
  15 & NGC 3998 & 270   & \cite{Francesco2006}  & 406 & 400 & 304\\
  16 & NGC 4051 & 1.73  & \cite{Denney2010}       & 120 & 160 & 84\\
  17 & NGC 4151 & 13.3  & \cite{Peterson2004} & 280 & - & -\\
  18 & NGC 4258 & 39    & \cite{Pastorini2007,Ferrarese2005} & 233 & 194 & 134\\
  19 & NGC 4303 & 5     & \cite{Pastorini2007}  & 160 & 160 & 109\\
  20 & NGC 4342 & 330   & \cite{Ferrarese2005} & 210 & - & 251\\
  21 & NGC 4395 & 0.36  & \cite{Peterson2005} & 40  & 90 & 90\\
  22 & NGC 4435 & <7.5  & \cite{Coccato2006}        & 160 & - & 157\\
  23 & NGC 4593 & 15    & \cite{Peterson2004} & - & - & 198\\
  24 & NGC 5128 & 49    & \cite{Graham2008}       & 250 & 170 & 120\\
  25 & MW (BH)  & 3.7   & \cite{Graham2008}       & 220 & 230 & -\\
  26 & MW (NUCL)    & 30    & \cite{Graham2009}     & 220 & 230 & -\\
  27 & Circunus & 1.7   & \cite{Gultekin2009}       & 130 & 152 & -\\
  28 & IC342 (BH)   & <0.5  & \cite{Boker1999}        & 80  & 218 & 74\\
  29 & IC342 (NUCL) & 6 & \cite{Boker1999}        & 80  & 218 & 74\\
  30 & 3C 120   & 30    & \cite{Ferrarese2005} & 100 & 280 & 100\\
  31 & MRK 79   & 52.4  & \cite{Peterson2004} & - & 150 & 130 \\
  32 & MRK 279  & 34.9  & \cite{Peterson2004} & 90 & 200 & -\\
  33 & NGC 428 (NUCL)& 3.16 & \cite{Seth2008}       & 40  & 110 & 30\\
  34 & NGC 524  & 830   & \cite{Krajnovic2009} & 290 & 320 & 250\\
  35 & NGC 2787 & 41    & \cite{Graham2008}       & 170 & 220 & 200\\
  36 & NGC 3245     & 210   & \cite{Graham2008}       & 150 & 200 & 210\\
  37 & NGC 3516     & 31.7  & \cite{Denney2010}       & 115 & 180 & 150\\
  38 & NGC 7457     & 3.5   & \cite{Ferrarese2005} & 58  & 130 & 65\\
  39 & NGC 7469 & 12.2  & \cite{Peterson2004} & 100 & 120 & 130\\
\hline
\end{tabular}
} \label{tabl1}
\end{table}


\begin{figure}
  \centerline{
  \subfloat[]{\label{1a-Mbh-V1-BTA}\includegraphics[width=0.5\textwidth]{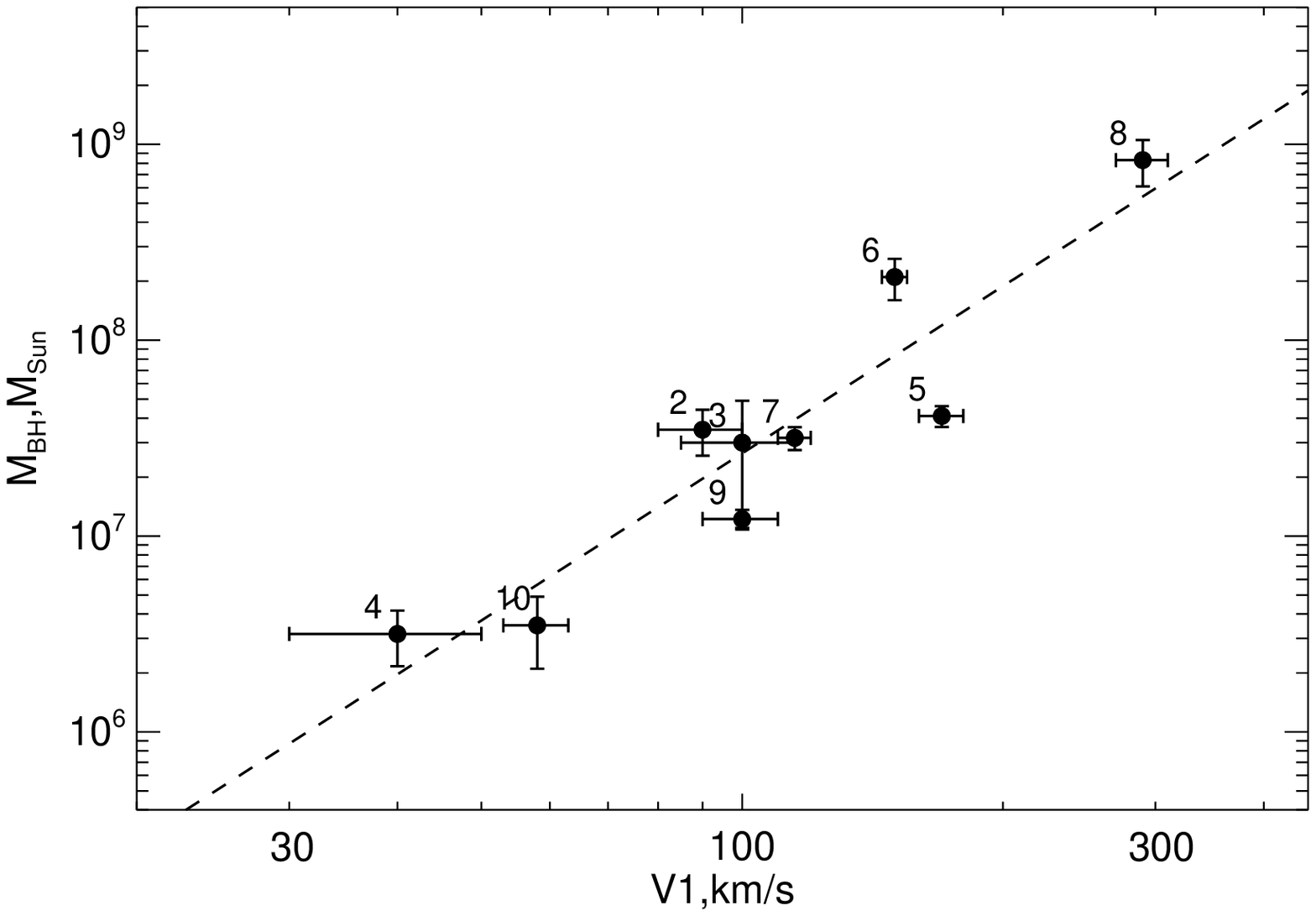}}
  \subfloat[]{\label{1b-Mbh-Vfar-BTA}\includegraphics[width=0.5\textwidth]{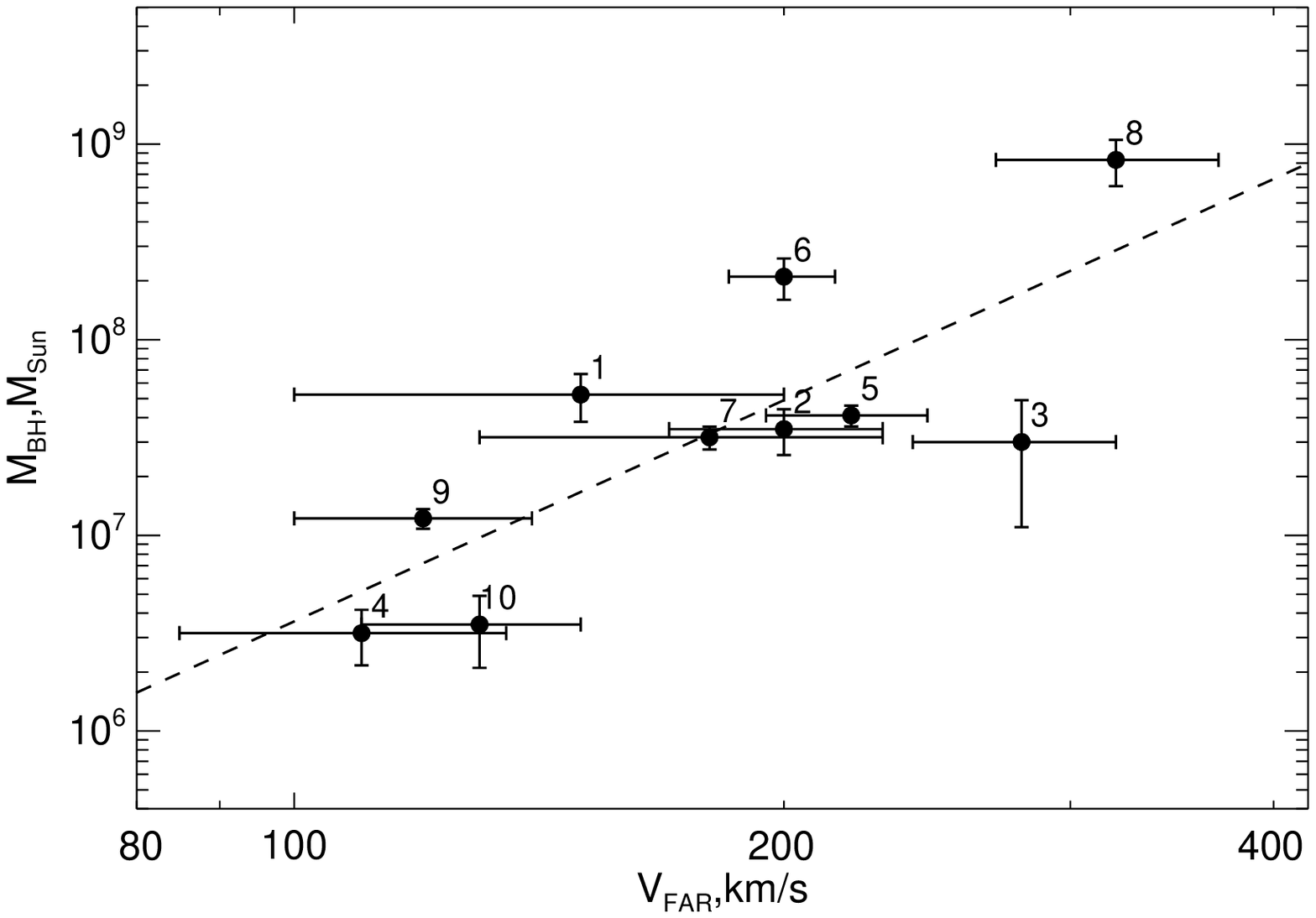}}}
  \caption{Relationship between the SMBH mass (the NC mass for NGC 428) and (a) the circular
  velocity at $R \approx$ 1 kpc (left) or (b)
the asymptotic rotational velocity. The numbers denote: 1 Mrk 79, 2
Mrk 279, 3 3C 120, 4 NGC 428, 5 NGC 2787, 6 NGC 3245, 7 NGC 3516, 8
NGC 524, 9 NGC 7469, 10 NGC 7457. There is no estimate of $V1$ for
Mrk 79.}
  \label{fig_1_BTA}
\end{figure}

\begin{figure}
\centerline{
\includegraphics[width=0.6\textwidth]{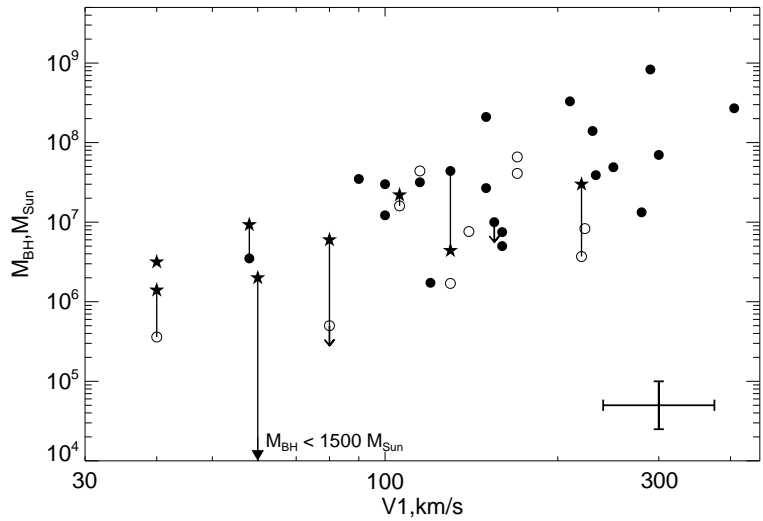}
} \caption{A comparison of masses of SMBHs (filled circles) and several NCs (asterisks) as
with the circular velocity of parent galaxies at $R \approx$ 1 kpc. The vertical lines
connect the SMBH and NC masses for the same galaxy. The upper limit on the SMBH mass
forM33 is indicated. The empty circles correspond to galaxies with pseudobulges.}
\label{2-Mbh-V1-general}
\end{figure}

\begin{figure}
  \centerline{
  \subfloat[]{\label{3a-Mbh-Vfar_general}\includegraphics[width=0.5\textwidth]{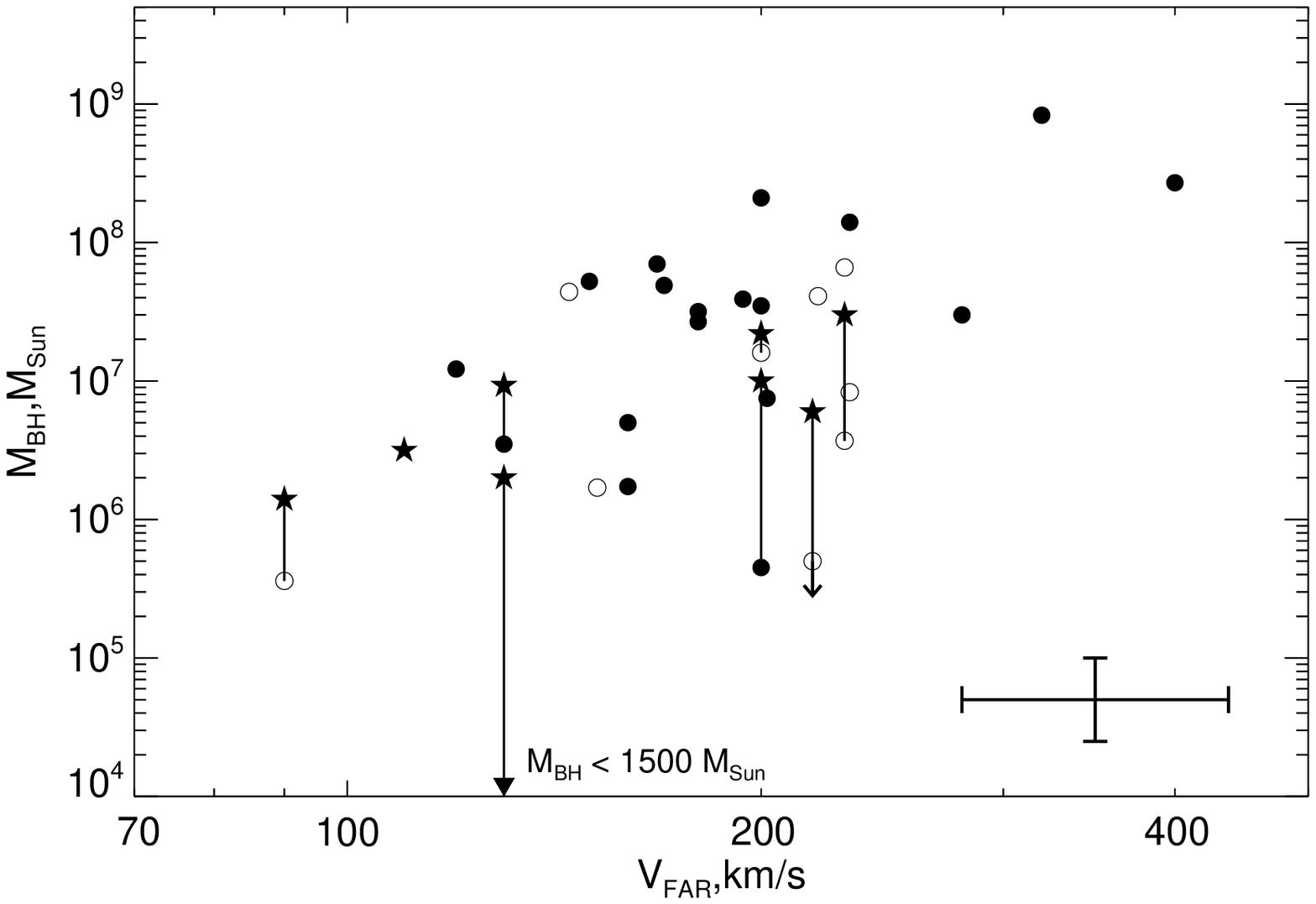}}
  \subfloat[]{\label{3b-Mbh-v2r}\includegraphics[width=0.5\textwidth]{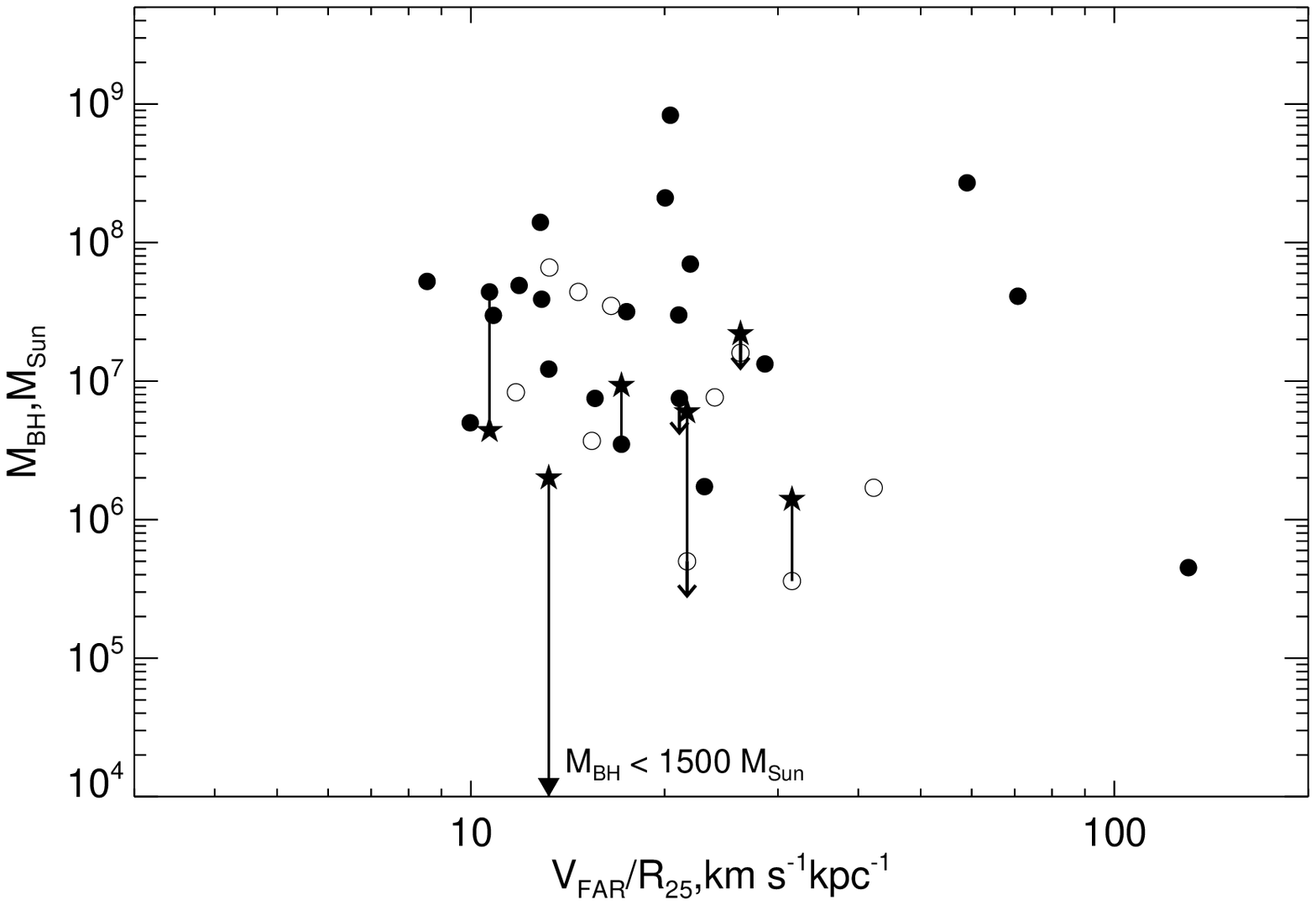}}}
  \centerline{
  \subfloat[]{\label{3c-Mbh-sigm_general}\includegraphics[width=0.5\textwidth]{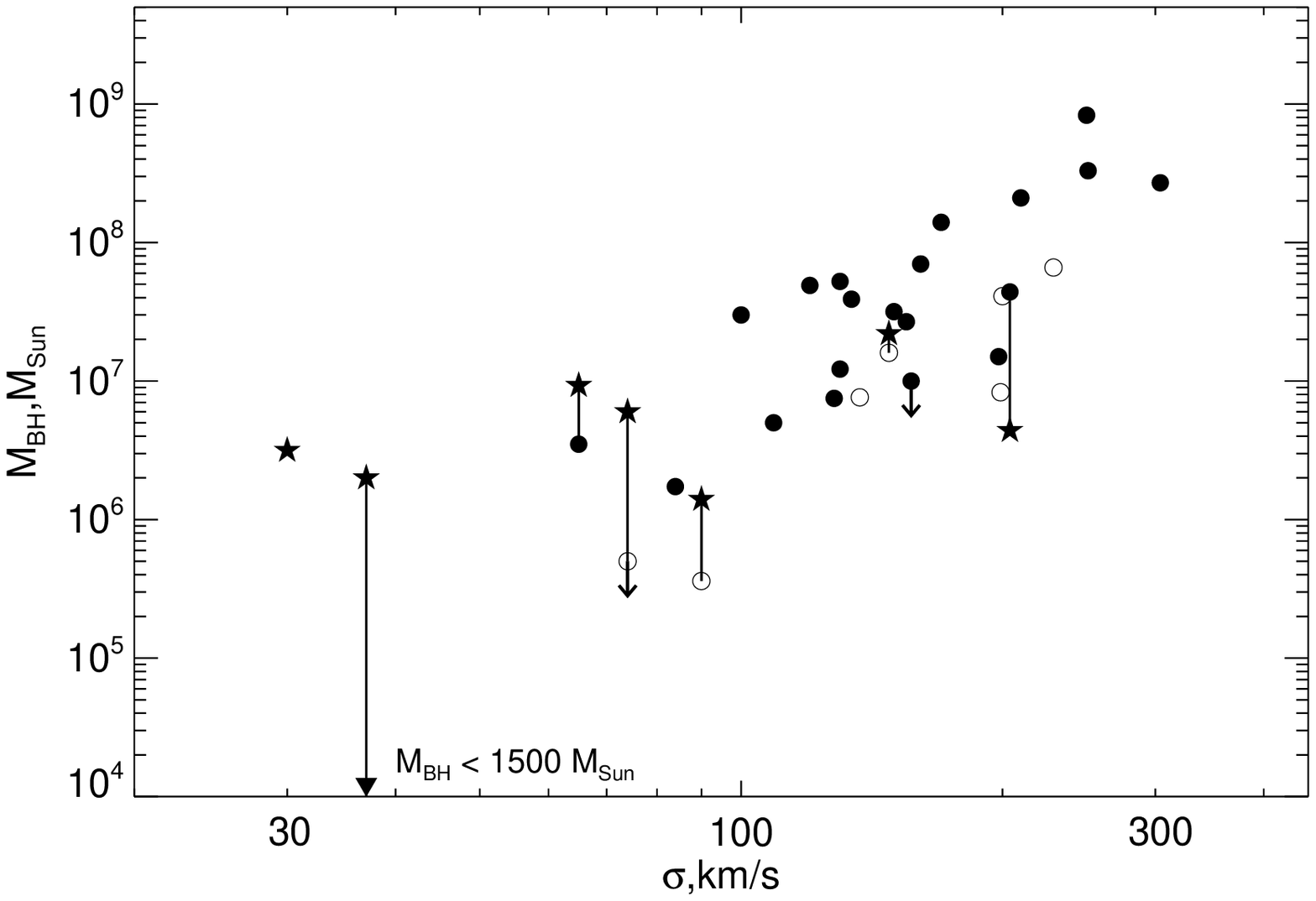}}
   }
  \caption{A  plot of the masses of SMBHs and several
NCs against (a) the circular velocity of parent galaxies far from the center $V_{FAR}$,
(b) the angular velocity of the galaxy at the optical radius $R = R_{25}$, and (c) the
central velocity dispersion taken from HYPERLEDA database. The notation is the same as in
Fig. 2.}
  \label{fig3-fulldata}
\end{figure}

\begin{figure}
  \centerline{
  \subfloat[]{\label{4a-Mbh-logv}\includegraphics[width=0.5\textwidth]{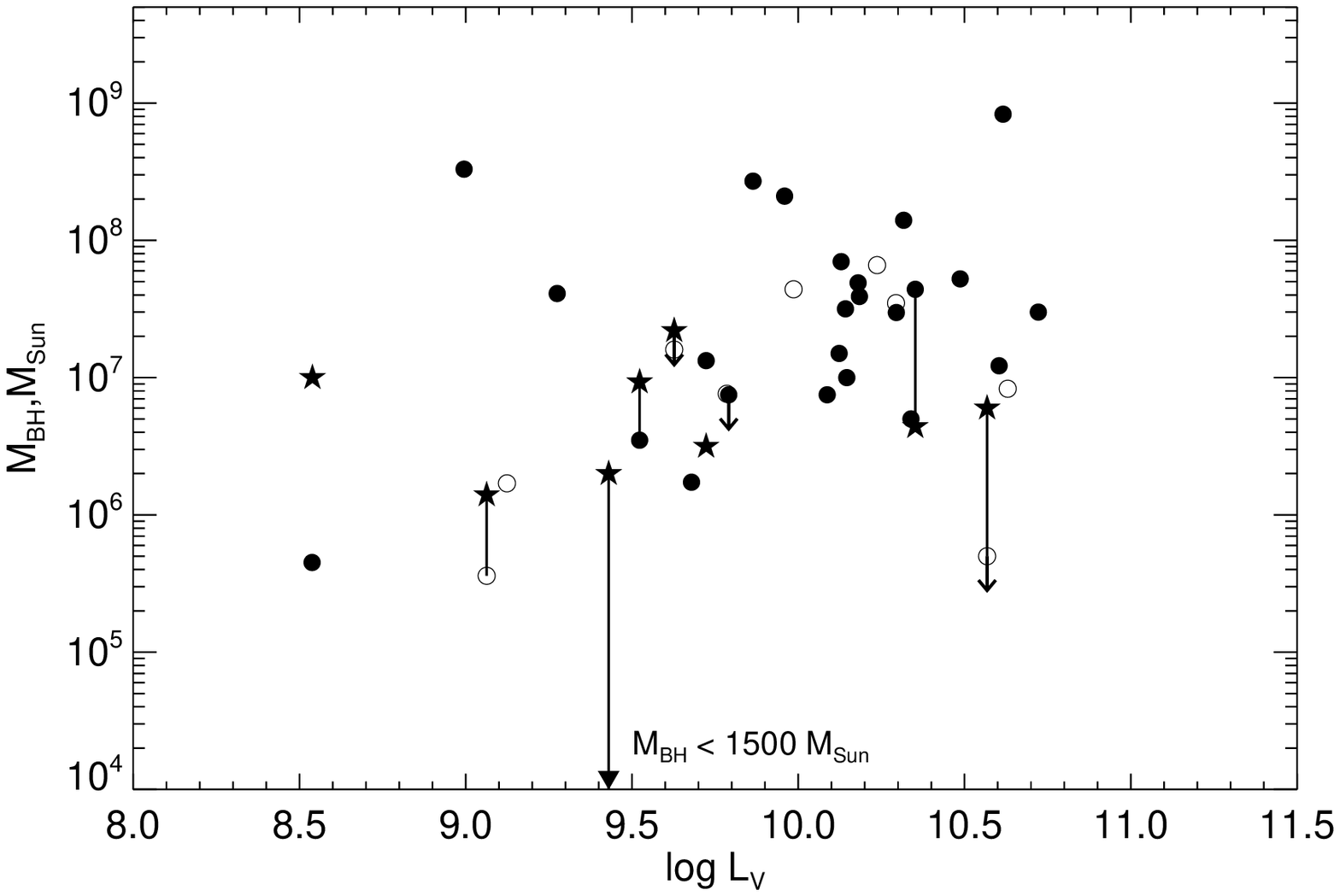}}
  \subfloat[]{\label{4b-Mbh-M25}\includegraphics[width=0.5\textwidth]{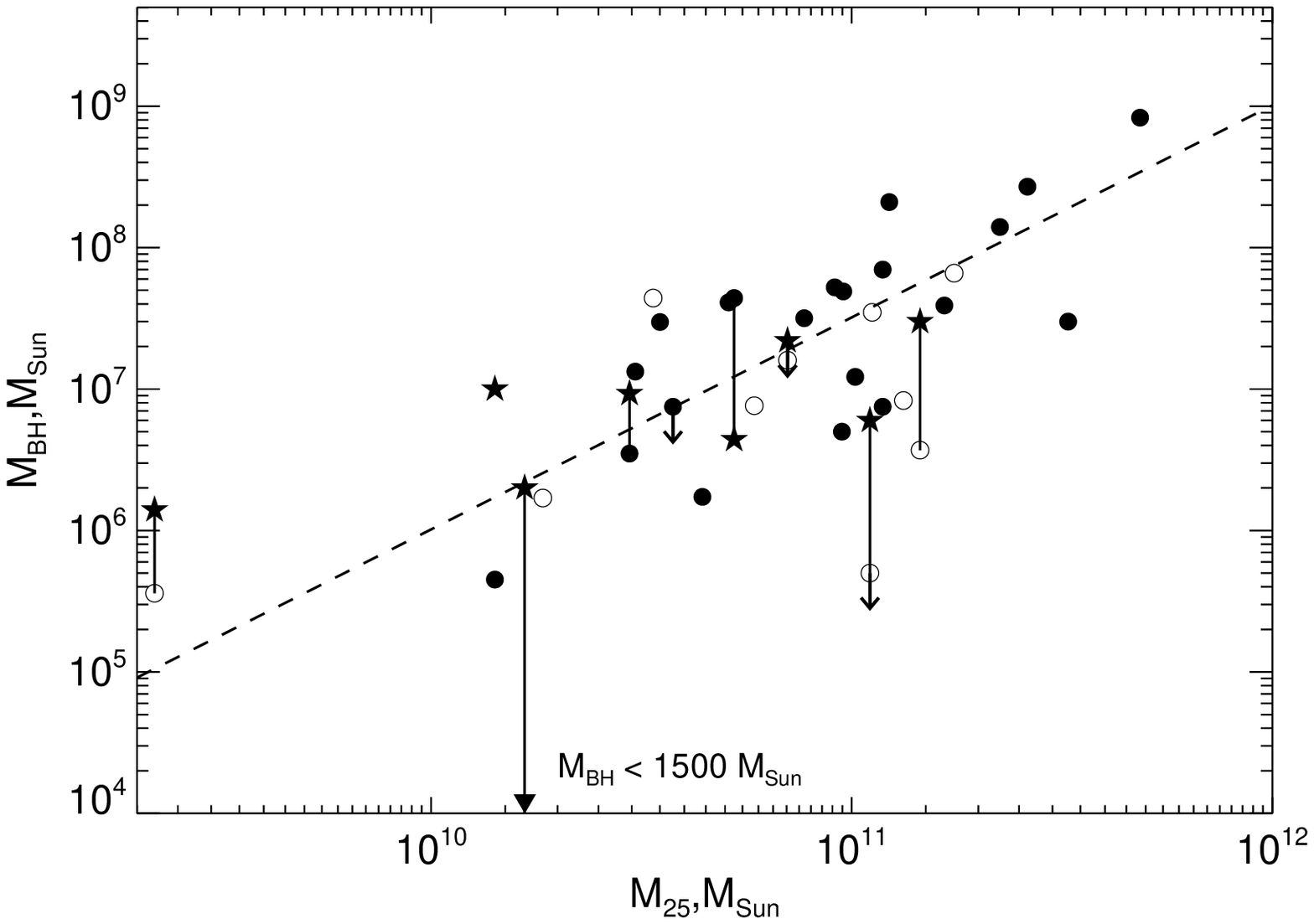}}
  }
  \caption{A comparison of the SMBH masses with (a) the total luminosities of parent  galaxies
  $L_V$ and (b) the dynamical (indicative) masses
within the optical radius $M_{25}=V^2_{far}R_{25}/G$. The notation is the same as in Fig.
2.}
  \label{fig4_logv_M25}
\end{figure}

\begin{figure}
  \centerline{
  \subfloat[]{\label{5a-Mnuc_Vrot}\includegraphics[width=0.5\textwidth]{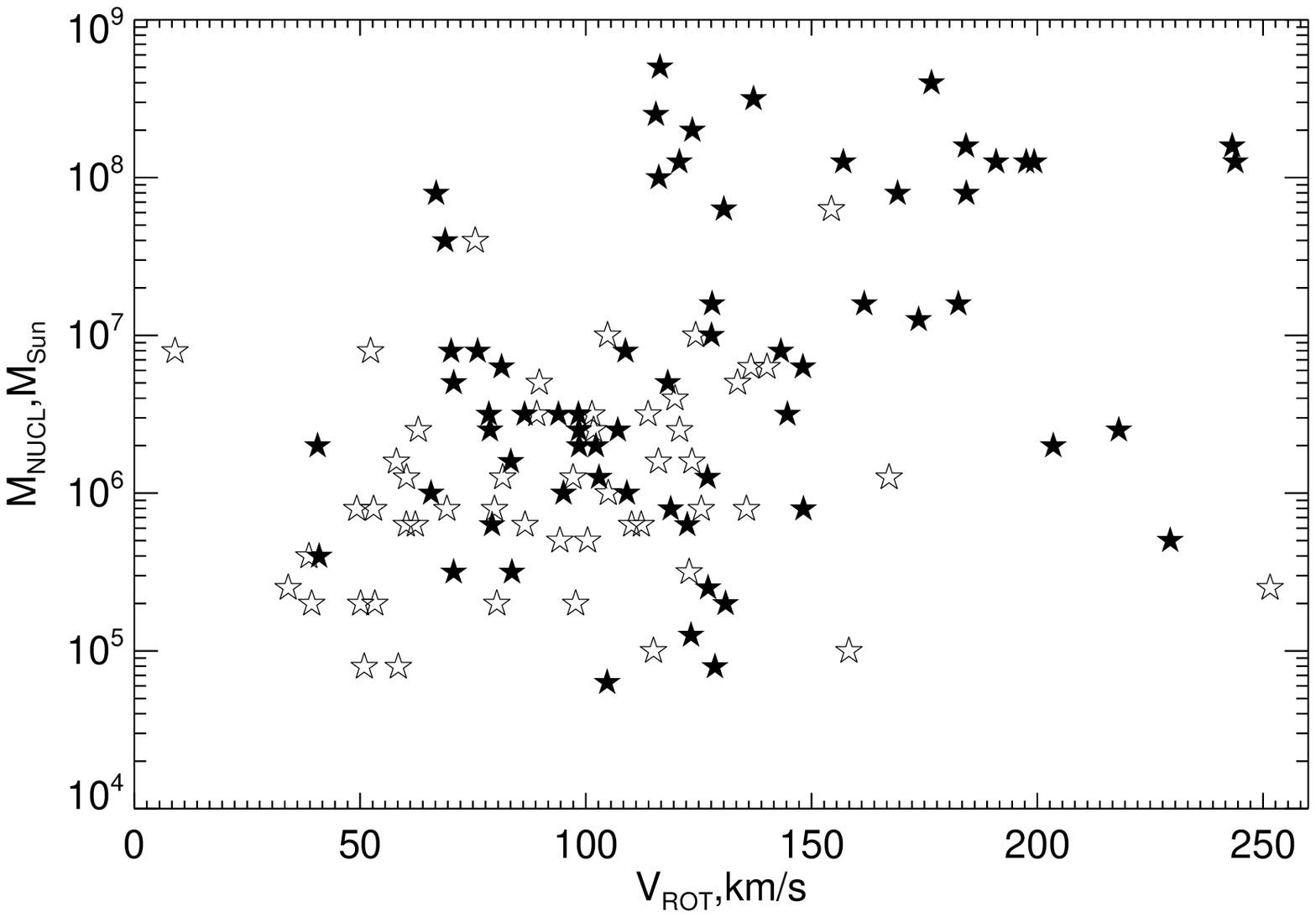}}
  \subfloat[]{\label{5b-Mnuc_Mg}\includegraphics[width=0.5\textwidth]{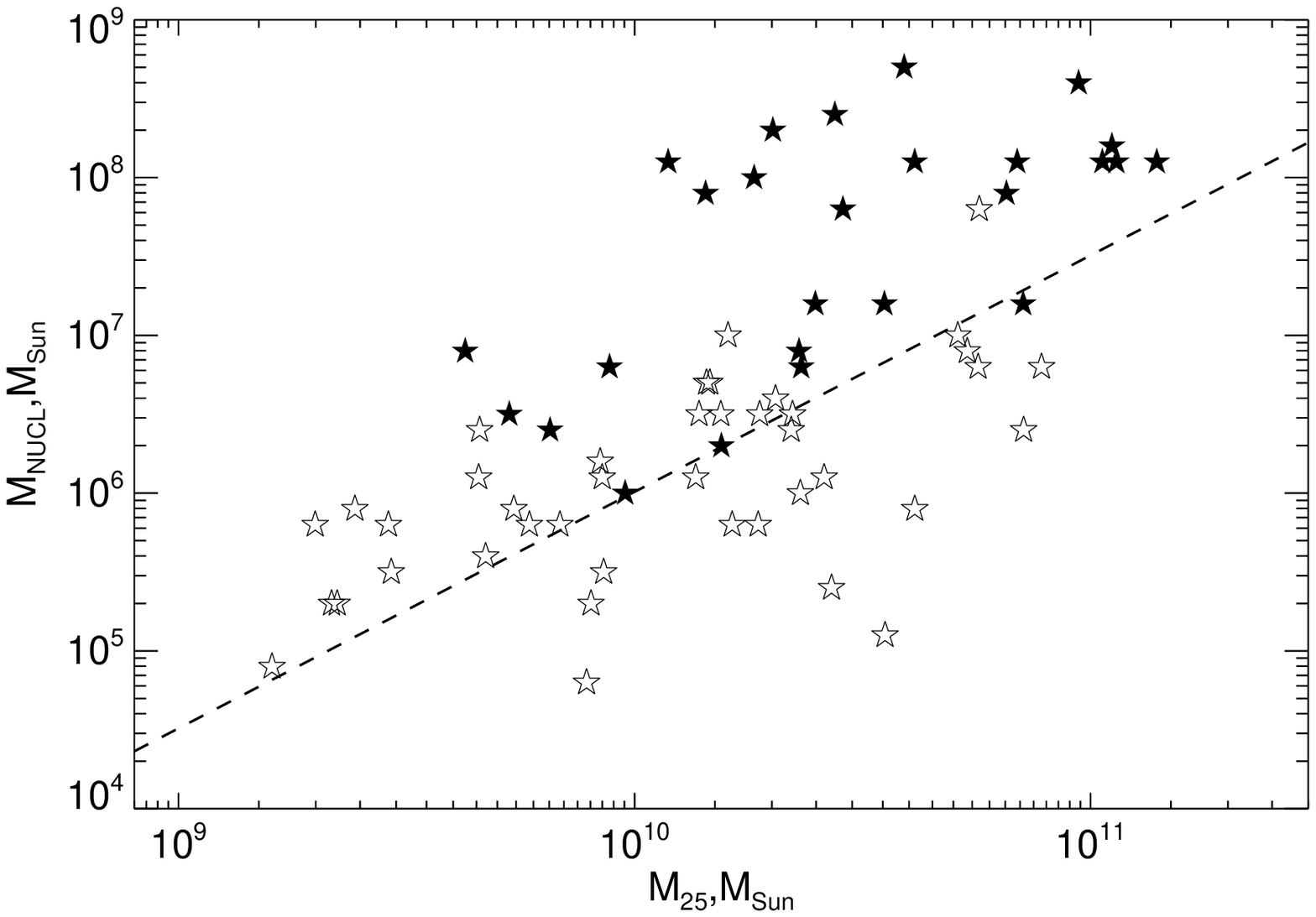}}
  }
  \caption{(a) Diagram illustrating the absence of a correlation between the NC masses
  (taken from [16]) and the velocity of rotation
(taken from HYPERLEDA). The filled symbols show S0–Sbc galaxies and
the empty symbols are for later-type galaxies. (b) Relation between
the masses of the NCs and the dynamical masses $M_{25}$ of parent
galaxies. A comparison with Fig.4b shows that the nuclear clusters
in S0-Sbc galaxies have higher (in the mean) masses of NCs than
SMBHs for a given $M_{25}$. The opposite is true for massive
late-type galaxies. }
  \label{nucl}
\end{figure}

\end{document}